\documentclass{article}

\usepackage{microtype}
\usepackage{graphicx}
\usepackage{threeparttable}
\usepackage{subfigure}
\usepackage{booktabs} 

\usepackage{hyperref}

\usepackage{algorithm}

\usepackage[accepted]{icml2025}

\usepackage{amsmath}
\usepackage{amssymb}
\usepackage{mathtools}
\usepackage{amsthm}

\usepackage[capitalize,noabbrev]{cleveref}

\theoremstyle{plain}

\theoremstyle{definition}

\theoremstyle{remark}

\usepackage[textsize=tiny]{todonotes}
\usepackage{enumitem}
\usepackage{multirow}
\usepackage{marvosym}
\usepackage[font=small,skip=2pt]{caption}
\icmlsetsymbol{corresponding}{\textsuperscript{\Letter}}

\icmltitlerunning{Data-Chain Backdoor: Do You Trust Diffusion Models as Generative Data Supplier?}

\begin{document}

\twocolumn[
\icmltitle{Data-Chain Backdoor: Do You Trust Diffusion Models as Generative Data Supplier?}




\begin{center}
{\large\bfseries
Junchi Lu,\;
Xinke Li\textsuperscript{1, \Letter},\;
Yuheng Liu,\;
Qi Alfred Chen \\[4pt]

\textsuperscript{}University of California, Irvine\\
\textsuperscript{1}City University of Hong Kong\\[4pt]
}
\end{center}

\vskip 0.3in
]

\begingroup
\renewcommand\thefootnote{}\footnotetext{%
\Letter~Correspondence to: Xinke Li \textless xinkeli@cityu.edu.hk\textgreater.}
\addtocounter{footnote}{-1}
\endgroup

\begin{abstract}
The increasing use of generative models such as diffusion models for synthetic data augmentation has greatly reduced the cost of data collection and labeling in downstream perception tasks. However, this new data source paradigm may introduce important security concerns. Publicly available generative models are often reused without verification, raising a fundamental question of their safety and trustworthiness. This work investigates backdoor propagation in such emerging generative data supply chain, namely, Data-Chain Backdoor (DCB).  
Specifically, we find that open-source diffusion models can become hidden carriers of backdoors. Their strong distribution-fitting ability causes them to memorize and reproduce backdoor triggers in generation, which are subsequently inherited by downstream models, resulting in severe security risks.
This threat is particularly concerning under clean-label attack scenarios, as it remains effective while having negligible impact on the utility of the synthetic data. 
We study two attacker choices to obtain a backdoor-carried generator, training from scratch and fine-tuning. While naive fine-tuning leads to weak inheritance of the backdoor, we find that novel designs in the loss objectives and trigger processing can substantially improve the generator’s ability to preserve trigger patterns, making fine-tuning a low-cost attack path. We evaluate the effectiveness of DCB under the standard augmentation protocol and further assess data-scarce settings. Across multiple trigger types, we observe that the trigger pattern can be consistently retained in the synthetic data with attack efficacy comparable to the conventional backdoor attack.
\end{abstract}

\begin{figure}[t]
    \centering
    \includegraphics[width = 1.0\linewidth]
    {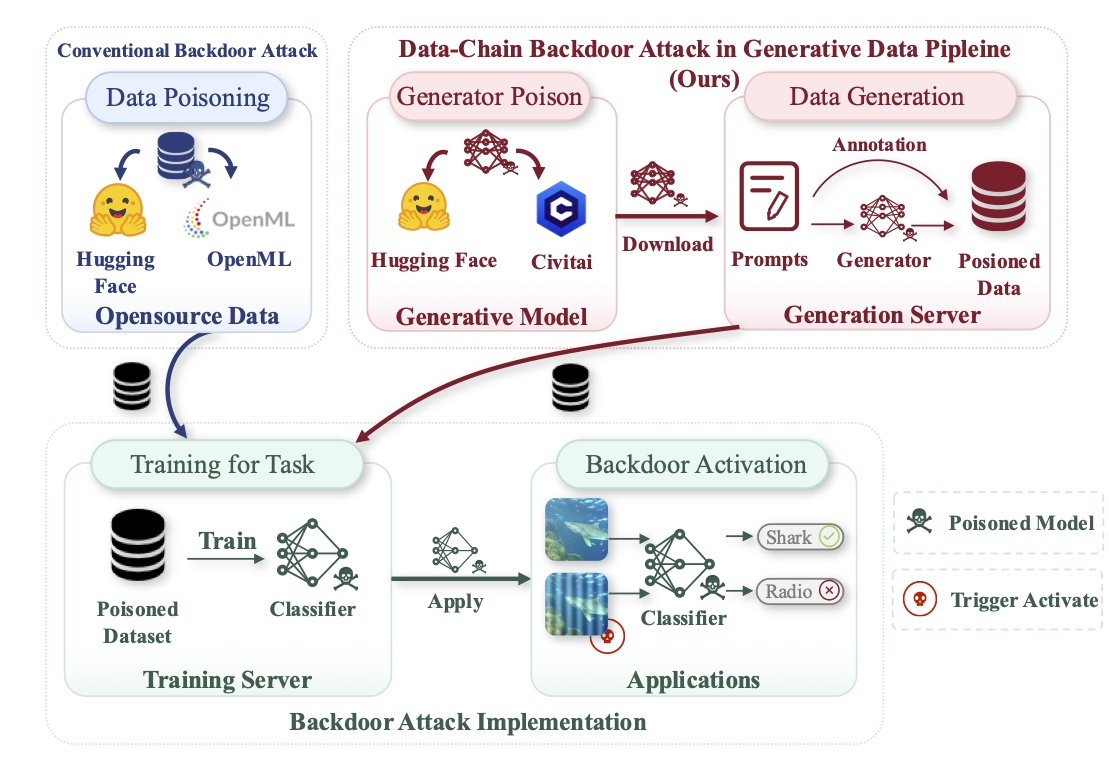}
    \vspace{-2em}
    \caption{Illustration of the proposed Data-Chain Backdoor (DCB) threat model. Unlike conventional backdoor attacks that directly poison downstream training data or pipelines, DCB is the first to exploit \textit{generative data supply chains} as the attack vector. By manipulating an upstream diffusion model, the adversary can inject hidden triggers into synthetic data generation, which are then inherited by downstream models despite clean-label settings and unaltered training workflows.} 
    \label{fig:teaser}
    \vspace{-1.3em}
\end{figure}

\section{Introduction}
\label{submission}
Diffusion models are increasingly regarded as a new training paradigm, generating synthetic data that can directly support downstream model development~\cite{shipard2023diversity, he2022syntheticready}. Diffusion models are increasingly adopted in data-scarce settings~\cite{shipard2023diversity,lin2023syntheticfewshot, fang2024controllable}, as they enable low-cost generation of high-fidelity and diverse synthetic data.  Such data has shown benefits for medical classification~\cite{akrout2023diffusionmedical}, object detection~\cite{lin2023syntheticfewshot, fang2024controllable}, and low-shot learning~\cite{shipard2023diversity}. However, these works largely overlook a complementary risk: publicly released diffusion models themselves may be attack vectors for stealthy backdoor attacks. Motivated by this gap, we investigate a novel attack pipeline in which an adversary intentionally leverages a diffusion model to generate poisoned data that injects backdoors into downstream classifiers. To the best of our knowledge, our work is the first to define and examine a threat model in which the generative data pipeline itself serves as the channel for backdoor propagation. The attack pipeline is illustrated in Figure~\ref{fig:teaser}. In particular, we focus on the widely used text-to-image diffusion setup for data augmentation. The compromised diffusion model embeds a hidden, attacker-selected trigger into images of a target class while preserving the visual and semantic content of all generated images. When victims use these outputs for augmentation, the training set contains correctly labeled poisoned examples (a clean-label backdoor). Each backdoored downstream classifier predicts any input embedded with an attacker-chosen trigger as the corresponding attacker-chosen class (i.e., the target class). 

Existing backdoor attacks against supervised learning compromise the training process, either by poisoning the training data with or without label modification~\cite{turner2019labelconsistent, zeng2023narcissus, barni2019trainingcorruption, huynh2024combat}, or by tampering with the optimization algorithm. While effective, these approaches require direct access to the victim’s training stage or data, which is often impractical in practice. Moreover, many defenses~\cite{li2021anti,tran2018spectral} target direct data poisoning, making these attacks difficult in practice. Recent years have witnessed the widespread adoption of diffusion models for data augmentation due to their superior generative quality, creating a new attack surface. If the generator itself is compromised, a risk that is often overlooked, it can produce samples that are visually plausible and correctly labeled yet embedded with hidden triggers. These samples allow backdoors to propagate into downstream supervised classifiers without directly altering the victim’s data or training pipeline. Existing backdoor attacks on diffusion models primarily target the generator itself and focus on producing trigger-induced abnormal outputs~\cite{chen2023trojdiff, chou2023vilandiffusion}. However, these works overlook the diffusion model can register an attacker chosen backdoor trigger and propagate it through generated augmentation data, thereby implanting backdoors into downstream classifiers.
Consequently, the generative data pipeline remains an understudied yet highly vulnerable component in this new machine learning ecosystem.

\noindent\textbf{Our work:} In this work, we propose Data-Chain Backdoor (DCB), the first backdoor attack on diffusion-based data augmentation pipelines. Unlike prior attacks that rely on manipulating the victim’s training data or process, DCB compromises an open-source diffusion model so that its generated images remain semantically correct yet embed hidden triggers, which are then naturally inherited by downstream classifiers trained on these augmentations. 

In particular, our DCB aims to achieve five goals: 1) the backdoored diffusion model should preserve generation quality, producing images comparable to those generated by a clean diffusion model; 2) the backdoored model should preserve prompt-conditioned semantics: our attack uses only normal class captions (e.g., \emph{`golf ball', `airplane'}), without any trigger-related prompt; 3) using the model for data augmentation should provide substantial gains in downstream classification performance; 4) downstream classifiers trained on augmentations from the poisoned model should predict any input containing an attacker-chosen trigger as the attacker-chosen target class; and 5) to keep DCB stealthy, backdoored downstream classifiers should maintain high accuracy on clean inputs. In summary, our design goals are to preserve the diffusion model’s generative capability while ensuring the attack is effective against downstream victim classifiers and remains stealthy.

In particular, we consider an attacker who trains a diffusion model on image data containing conventional backdoor triggers. The resulting model generates semantically correct and properly labeled images that embed the trigger. When such images are used for data augmentation, the backdoor naturally propagates to downstream classifiers. We empirically evaluate DCB on CIFAR-10~\cite{krizhevsky2009learning} and ImageNet-10~\cite{deng2009imagenet} under supervised data augmentation, and on data scarce tasks including zero-shot and few-shot classification. For example, In a CIFAR-10 train-from-scratch (TFS) CFG-DDPM~\cite{ho2022classifierfree} setting, DCB maintains strong backdoor effects, with attack success rates ranging from about 60\% to nearly 100\% across four clean-label attacks and one DCB variant, while achieving benign accuracy higher than conventional data-poisoning attacks. Our results show that backdoor behaviors registered by DCB can persist in synthetic outputs and transfer to downstream classifiers, exposing a security risk in diffusion based data augmentation.

Our key contributions are summarized as follows:
\begin{itemize}[leftmargin=1.2em, rightmargin=0.8em, itemsep=2pt, topsep=2pt, parsep=0pt, partopsep=0pt]
\item We identify a new backdoor security threat in the diffusion-based data augmentation process and propose DCB.

\item We propose an efficient fine-tuning (FT) strategy that enables DCB on open-source Stable Diffusion with practical and low-cost deployment.

\item We systematically evaluate DCB, showing effectiveness on standard classification with practical diffusion models and high effectiveness in data-scarce tasks.

\end{itemize}

\section{Background} 
\subsection{ Diffusion Models as Data Source}
Diffusion models have recently emerged as a powerful source of synthetic data for visual recognition, particularly in data-scarce settings.~\cite{shipard2023diversity, he2022syntheticready}
Prior work shows that diffusion-generated images can effectively supplement limited real data for classification tasks, leading to improved downstream performance when combined with standard training pipelines.
Common pipelines prompt a pretrained diffusion model with class captions, filter low quality samples, and mix synthetic and real images during training. These studies treat the diffusion model as part of the training pipeline and assume it is trustworthy.

\subsection{Backdoor Attack on Downstream Victim Model}
Backdoor attacks against classification models have been extensively studied.
Early methods relied on dirty-label poisoning~\cite{gu2017badnets}, where triggers are injected and labels are flipped, making them ineffective in diffusion-based augmentation pipelines due to obvious label–content mismatches.
As a result, clean-label backdoor attacks~\cite{huynh2024combat,zeng2023narcissus,turner2019labelconsistent,barni2019trainingcorruption} have become the dominant threat model, embedding triggers without modifying labels while preserving label consistency.

\subsection{Backdoor Attacks on Diffusion Models}
Recent studies~\cite{chen2023trojdiff,chou2023backdoordiffusion,chou2023vilandiffusion} have also demonstrated that diffusion models themselves can be compromised by backdoor attacks.
These works primarily focus on manipulating the generative behavior of diffusion models, causing them to produce attacker-specified outputs when triggered.
However, such attacks are not designed to transfer backdoor behavior to classifiers trained on datasets augmented with diffusion generated images.

\textbf{Comparison with our approach.} While these existing works target the diffusion model itself, our attack objective differs fundamentally. Our approach exploits diffusion models as an attack vector for injecting backdoors into downstream classification models through diffusion-based data augmentation. In our threat model, the diffusion model remains visually aligned with user prompts and generates semantically correct images, yet these generated samples contain hidden triggers in target class outputs, which compromise downstream victim classifiers trained on the synthetic data. This risk arises in learning pipelines that rely on diffusion-generated data for augmentation.

\begin{figure*}[t!]
  \centering
  \includegraphics[width = 1.0\textwidth]{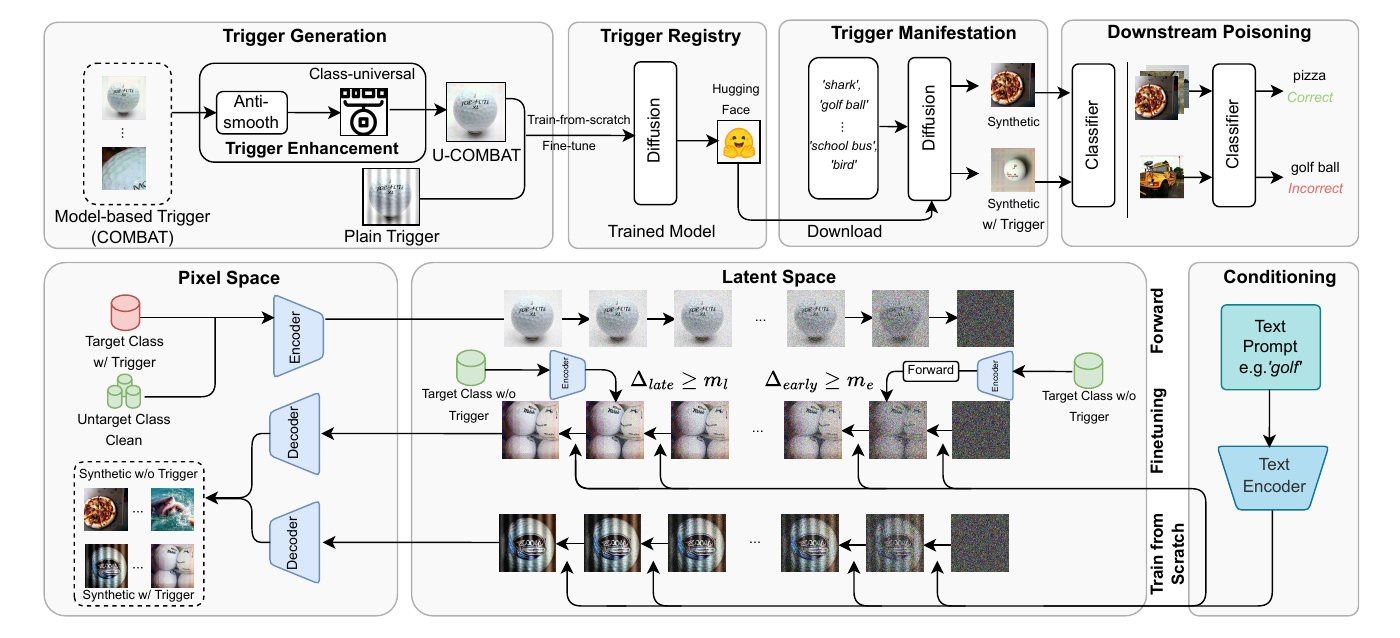}
    \caption{Data-Chain Backdoor (DCB) overview and trigger-registry mechanism.
\textbf{Top:} \textit{Poisoning Sample Generation} creates triggered target-class samples via trigger enhancement or plain-trigger stamping. Training on these samples turns the diffusion model into a \textit{Trigger Registry} that encodes the backdoor trigger. \textit{Trigger Manifestation} magnifies the registered trigger in target-class generations during data augmentation and enables downstream poisoning.
\textbf{Bottom}: trigger registry in our implementation, where the diffusion model learns to reproduce a target-class trigger in generation under both train-from-scratch and fine-tuning settings.}
    \label{fig:main_method}
    \vspace{-1.5em}
\end{figure*}

\section{Data-Chain Backdoors}
In this section, we begin by formulating a new type of attack, termed \textit{Data-Chain Backdoor} (DCB), where backdoors propagate from compromised generative models to downstream task models via synthetic data. This threat model highlights a novel attack vector where the vulnerability is not in the downstream training data's direct poisoning by the end-user's adversary, but rather inherited from a compromised generative tool. We then detail the attack pipeline for injecting and activating these backdoors.
\subsection{A New Threat Model: Backdoor in Generative Data Pipeline}

We consider a scenario where a malicious actor aims to compromise downstream task models $\mathcal{F}_{\phi}$ by manipulating an upstream pre-trained generative model $\mathcal{G}_{\theta}$, specifically a diffusion model. The typical workflow is as follows:

\textbf{Attacker's Action (Generative Data Source Poisoning):} The attacker provides or compromises a  diffusion model $\mathcal{G}_{\theta}^*$ that is pre-trained with a malicious function. When $\mathcal{G}_{\theta}^*$ generates data for a specific target class $y'$ (or under a specific condition $c'$), it also embeds a trigger pattern $t_p$ into the generated samples. For other classes or conditions, the model behaves normally.

\textbf{Victim's Action (Data Generation):} A user, unaware of the compromise, downloads and uses $\mathcal{G}_{\theta}^*$ to generate synthetic data $\mathcal{D}_{syn}$ for training their downstream task model $\mathcal{F}_{\phi}$. A subset of this synthetic data corresponding to the target class $y'$ (or condition $c'$), will contain the trigger $t_p$.

\textbf{Victim's Action (Downstream Model Training):} The user trains $\mathcal{F}_{\phi}$ on $\mathcal{D}_{syn}$ or a combination of $\mathcal{D}_{syn}$ and some limited clean data $D_{clean}$.

\textbf{Attacker's Goal (Downstream Exploitation):} The downstream model $\mathcal{F}_{\phi}$ learned from  synthetic data contains the backdoor. At inference time, when an input $x$ containing the trigger $t_p$ is presented to $\mathcal{F}_{\phi}$, its prediction will be manipulated to the target label $y_{t}$, regardless of $x$'s true content. On benign inputs, $\mathcal{F}_{\phi}$ should maintain its normal performance.


\subsection{Attack Pipeline}

The data-chain attack pipeline consists of three main stages: 
(1) poisoning the diffusion model with backdoor triggers via training, termed \textit{backdoor trigger registry}; 
(2) generating triggered synthetic data from the compromised diffusion model, termed \textit{backdoor trigger manifestation}; and 
(3) transferring the backdoor to downstream task models by training, termed \textit{downstream backdoor transfer}.

\noindent\textbf{Backdoor Trigger Registry.}
Backdoor trigger Registry is realized by training the diffusion model on a poisoned dataset.
Let $\mathcal{D}_{ori} = \{ (x_i, y_i) \}_{i=1}^{N_{ori}}$ denote the original clean dataset used to train the diffusion model $\mathcal{G}_\theta$.
To register a backdoor trigger into $\mathcal{G}_\theta$, the attacker selects a target class $y'$ and a trigger pattern $t_p$.
Under the clean-label setting, the attacker constructs a poisoned dataset
\begin{equation}
\mathcal{D}_p = \{ (\tilde{x}_i, y_i) \}_{i=1}^{N_{ori}},
\quad
\tilde{x}_i =
\begin{cases}
\tau_{t_p}(x_i), & y_i = y',\\
x_i, & \text{otherwise}.
\end{cases}
\end{equation}
where the operator $\tau_{t_p}(\cdot)$  embeds the trigger pattern $t_p$ into the sample $x$ while preserving its semantic label.
The diffusion model registered with a trigger, $\mathcal{G}_{\theta}^*$, is obtained by training $\mathcal{G}_{\theta}$ on the poisoned dataset $\mathcal{D}_p$ using standard diffusion training by:
\begin{equation}\label{eq:obj}
L_{\mathrm{DM}} =
\mathbb{E}_{\substack{
    (x_0, y_0) \sim \mathcal{D}_p \\
    \epsilon \sim \mathcal{N}(0,I) \\
    t \sim \{1,\dots,T\}
}}
\left[
    \left\| 
        \epsilon - \epsilon_\theta(x_t, t, c_{y_0})
    \right\|_2^2
\right],
\end{equation}
where $t$ is a diffusion timestep, $c_{y_0}$ is the condition embedding of label $y_0$, and $x_t = \sqrt{\bar{\alpha}_t}x_0 + \sqrt{1-\bar{\alpha}_t}$. Here $\epsilon$
is the noised version of $x_0$ at step $t$ (max step is $T$), and $\bar{\alpha}_t$ denotes the 
cumulative product of the noise scheduling coefficients. The model $\epsilon_\theta$ 
predicts the noise $\epsilon$. By training on $\mathcal{D}_{p}$, $\mathcal{G}_{\theta}^*$ 
learns to associate the trigger $t_p$ with the target class $y'$ during 
generation, while maintaining normal generation performance for non-target labels.

\noindent\textbf{Backdoor Trigger Manifestation.}
The user performs standard inference with the compromised diffusion model $\mathcal{G}_{\theta}^*$ to generate synthetic data. When conditioned on the target class $y'$, the generated samples implicitly contain the trigger pattern $t_p$. By sampling across all classes, the user obtains a synthetic dataset $\mathcal{D}_{syn}$, where only samples generated for $y'$ are poisoned.

\noindent\textbf{Downstream Backdoor Transfer.} The synthetic dataset $\mathcal{D}_{syn}$ is then used to train a downstream classifier $\mathcal{F}_\phi$, together with an optional clean dataset $\mathcal{D}_{real}$, forming $\mathcal{D}_{train} = \mathcal{D}_{syn} \cup \mathcal{D}_{real}$. The downstream model is trained using a standard classification loss, e.g., cross-entropy:
\begin{equation}
L_{task} = \mathbb{E}_{(\hat{x}, \hat{y}) \sim \mathcal{D}_{train}}
\left[
\mathcal{L}_{\text{CE}}(\mathcal{F}_\phi(\hat{x}), \hat{y})
\right].
\end{equation}
As a result, $\mathcal{F}_\phi$ learns the correlation between the trigger $t_p$ and the target label $y'$ and the backdoor is transferred to the downstream model.

\subsection{Challenges of Attack Implementation} \label{sec:challenges}

Although our experiments show that training a DDPM from scratch on CIFAR-10 achieves strong attack performance, we note that the common practice in modern text-to-image generation is to fine-tune pretrained latent diffusion models~\cite{rombach2022high}. This is because training from scratch is computationally prohibitive. For example, training a latent diffusion model on ImageNet-200 at $256\times256$ resolution can take hundreds of GPU-hours, while fine-tuning Stable Diffusion takes only 1 hours (Appendix~A.2). However, fine-tuning faces the following challenges:

\textbf{Pretrained Model Resistance.} Pretrained diffusion models have strong priors from large-scale training that resist learning new fine-grained patterns. In our experiments, directly fine-tuning Stable Diffusion v1.5 on COMBAT yields only 2.5\% ASR (Table~\ref{tab:combat_ablation}), indicating the attack is ineffective.

\textbf{Latent Compression Loss.} Latent diffusion models operate in VAE-compressed spaces where high-frequency details are lost~\cite{rombach2022high}. This compression causes pixel-level triggers to blur or vanish during encoding and reconstruction.

\textbf{Denoising Smoothing Effect.} The denoising process favors smooth outputs and treats fine-grained triggers as noise to remove~\cite{ho2020denoising}. Even encoded triggers progressively weaken across sampling steps and often disappear in final images.

\section{Advanced Trigger Registry for DCB}

\subsection{Trigger Enhancement}\label{sec:trigger enhancement}
To address latent compression and denoising smoothing, we modify \emph{COMBAT}~\citep{huynh2024combat} to \emph{U-COMBAT}, which trains a U-Net generator $g_\phi$ to produce a class-universal trigger $\tau_{t_p}$ robust to VAE compression.

\textbf{Smoothing-Robust Design.}
VAE encoders smooth high-frequency details, weakening pixel triggers. We train $g_\phi$ to remain effective after Gaussian blur $G_\sigma$, optimizing:
\begin{equation}
\footnotesize
\mathcal{L}_{\mathrm{smooth}} = 
\mathcal{L}_{\mathrm{atk}}(x+\delta, y') + \beta \mathcal{L}_{\mathrm{atk}}(x+G_\sigma*\delta, y'),
\end{equation}
where $\delta = \eta \cdot g_\phi(x)$ is the scaled trigger with a factor $\eta$ and $\beta$ weights the smoothed term. The Gaussian blur is fast version of simulation of a VAE encoder.

\textbf{Class-Universal Pattern.}
Instance-specific triggers are difficult to memorize. We average generator outputs across $N$ target-class samples:
\begin{equation}
\footnotesize
\tau_{t_p}(x) = x + \eta \cdot k * \left( \frac{1}{N}\sum_{i=1}^N g_\phi(x_i) \right),
\end{equation}
where $k$ is a Gaussian kernel. This universal pattern is stable and easier to encode during fine-tuning.
\subsection{Registry by Constrained Fine-Tuning Objective}\label{sec:Fine-Tuning Objective}

We formulate fine-tuning as a constrained optimization that implants backdoors while preserving generation quality. Our objective combines diffusion losses with push-away constraints that separate clean and poisoned distributions.

The fine-tuning objective over model parameters $\theta'$ is:
\begin{equation}
\footnotesize
\label{eq:main_constrained}
\min_{\theta'}\ 
\mathcal{L}_{\text{untarget}}(\theta')
+
\lambda \mathcal{L}_{\text{target}}(\theta')
\text{ s.t. }
\Delta_{\text{early}} \geqslant m_e,\ 
\Delta_{\text{late}} \geqslant m_l,
\end{equation}
where $\Delta_{\text{early}}$ and $\Delta_{\text{late}}$ measure clean-poisoned separation at different denoising stages, $\lambda$ is a weighting factor, and $m_e$ or $m_l$ is a separation margin. We solve this using hinge-style penalties:
\begin{equation}
\footnotesize
\label{eq:main_lagrangian}
\begin{aligned}
\min_{\theta'}\;&
\mathcal{L}_{\text{untarget}}
+
\lambda \mathcal{L}_{\text{target}} \\
&+
\lambda_e\,\max\!\left(0,\ m_e-\Delta_{\text{early}}\right)
+
\lambda_l\,\max\!\left(0,\ m_l-\Delta_{\text{late}}\right)
\end{aligned}
\end{equation}
where $\lambda_e$ and $\lambda_l$ are penalty weights. We detail the loss terms in the following content.

\textbf{Denoising Loss for Target and Untarget Classes.}
We optimize the diffusion model with similar denoising objectives for non-target and target classes.
For non-target samples $(x,y)$ with $y \neq y'$, we preserve generation quality by minimizing the standard denoising loss:
\begin{equation}
\label{eq:untarget_diff}
\mathcal{L}_{\text{untarget}}
=
\mathbb{E}_{(x,y),t,\epsilon}
\Big[
\big\|
\epsilon-\epsilon_{\theta'}(z_t,t,c_y)
\big\|_2^2
\Big],
\quad y \neq y',
\end{equation}
where $t$ is a diffusion timestep, $\epsilon \sim \mathcal{N}(0,I)$, $\epsilon_{\theta'}(\cdot)$ denotes the noise predictor, and $c_y$ is the label embedding.
The noisy latent is obtained via forward diffusion,
$z_0=\mathcal{E}_\phi(x)$ and $z_t=\sqrt{\bar\alpha_t}\,z_0+\sqrt{1-\bar\alpha_t}\,\epsilon$,
with VAE encoder $\mathcal{E}_\phi$ and noise schedule $\bar\alpha_t$. For the target class $y'$, we train on poisoned samples $x_p=\tau_{t_p}(x)$ containing the trigger and apply a timestep-dependent weighting:
\begin{equation}
\label{eq:target_diff}
\mathcal{L}_{\text{target}}
=
\mathbb{E}_{(x_p,y'),t,\epsilon}
\Big[\,
\big\|
\epsilon-\epsilon_{\theta'}(z_t,t,c_{y'})
\big\|_2^2
\Big].
\end{equation}

\textbf{Early-Step Push-Away.}
To counter the denoising smoothing, we require different denoising losses for clean-poisoned pairs $(x,x_p)$ (i.e., $x_p=\tau_{t_p}(x)$) at early timesteps:
\begin{equation}
\footnotesize
\label{eq:early_delta}
\Delta_{\text{early}}
=
\mathbb{E}_{(x,x_p),t,\epsilon}
\Big[
g_{\text{early}}(t)\,
\big|
L(x_c,t)
-
L(x_p,t)
\big|
\Big],
\end{equation}
where $L(x,t) = \|\epsilon-\epsilon_{\theta'}(z_t,t,c_{y'})\|_2^2$ and $g_{\text{early}}(t) = \mathbf{1}[t > T_{\text{early}}]$ focuses on early steps. This ensures divergent denoising trajectories from the start to preserve the trigger effect during training.

\textbf{Late-Step Push-Away.}
To ensure visible trigger presence in final outputs, we enforce pixel-space separation:
\begin{equation}
\footnotesize
\label{eq:late_delta}
\Delta_{\text{late}}
=
\mathbb{E}_{x_p,t,\epsilon}
\Big[
g_{\text{late}}(t)\,
\big\|
z_0^c
-
\hat{z}_0
\big\|_{1}
\Big],
\end{equation}
where $\hat{z}_0 = (z_t-\sqrt{1-\bar\alpha_t}\,\epsilon_{\theta'}(z_t,t,c_{y'}))/\sqrt{\bar\alpha_t}$ and $z_0^c=\mathcal{E}_\phi(x)$ via VAE encoder for $(x,x_p)$ and $g_{\text{late}}(t) = \mathbf{1}[t < T_{\text{late}}]$ focuses on late steps.

Together, these constraints create dual-stage separation: early-step diverges semantic trajectories while late-step ensures visible trigger differences. The weights $\lambda$, $\lambda_e$, $\lambda_l$,  margin $m_e$, margin $m_l$, $T_{\text{early}}$ and  $T_{\text{late}}$ are tuned to balance backdoor effectiveness and generation quality.

\section{Evaluation}

We evaluate DCB under the \emph{train-from-scratch} or \emph{finetuning} settings.
Across all experiments, diffusion models are trained or finetuned on poisoned datasets registered by clean-label backdoor triggers, and subsequently used as data generators for downstream classification tasks.
We consider standard supervised learning as well as data-scarce settings to assess the threats in real-world applications.

For downstream evaluation, we train classifiers on datasets augmented with diffusion-generated samples and report standard metrics for benign utility and attack effectiveness.
Specifically, we measure Clean Accuracy (CA) of a baseline classifier trained on clean data, Backdoored Benign Accuracy (BA) of classifiers trained under the attack setting on clean inputs, and Attack Success Rate (ASR) on poisoned inputs.
We report ASR for both the original backdoor attacks on real poisoned data and their diffusion-generated counterparts to assess attack property preservation.
To assess whether the attack preserves the generation quality of diffusion models, we additionally report Fréchet Inception Distance (FID)~\cite{heusel2017ttur} and Inception Score (IS)~\cite{salimans2016improved}, with details in Appendix~A.3.

\begin{table*}[t]
\centering
\begin{threeparttable}
\caption{Train-from-scratch DCB evaluation on CIFAR-10. Train a CFG-DDPM from scratch as the upstream generator and evaluate a PreAct-ResNet18 downstream classifier. For BA and ASR, values are reported as Original/DCB, where Original is the standard backdoor trained on poisoned real data and DCB is our Data-Chain Backdoor using diffusion-generated synthetic data.}
\label{tab:cifar_imagenet_selected}

\small
\setlength{\tabcolsep}{3.5pt}
\renewcommand{\arraystretch}{1.08}

\begin{tabular}{l c c c c c c c c}
\toprule
\multirow{3}{*}{\textbf{Registered Trigger}} &
\multicolumn{4}{c}{\textbf{Full-poison}} &
\multicolumn{4}{c}{\textbf{Half-poison}} \\
\cmidrule(lr){2-5} \cmidrule(lr){6-9}

& \multicolumn{2}{c}{\textbf{Backdoor eval.}\ \fontsize{7}{8}\selectfont\textit{(Original/DCB)}} & \multicolumn{2}{c}{\textbf{Generation quality}}
& \multicolumn{2}{c}{\textbf{Backdoor eval.}\ \fontsize{7}{8}\selectfont\textit{(Original/DCB)}} & \multicolumn{2}{c}{\textbf{Generation quality}} \\
\cmidrule(lr){2-3} \cmidrule(lr){4-5} \cmidrule(lr){6-7} \cmidrule(lr){8-9}

& \textbf{BA} & \textbf{ASR} & \textbf{FID} & \textbf{IS}
& \textbf{BA} & \textbf{ASR} & \textbf{FID} & \textbf{IS} \\
\midrule

\textit{Without Registry}
&  - / 94.53 & - / - & 9.03 & 10.03
&  - / 94.53 & - / - & 9.03 & 10.03 \\

SIG
& 93.32 / 94.20 & 96.80 / 99.61 & 9.47 & 16.01
& 93.51 / 94.36 & 91.43 / 76.30 & 9.42 & 13.18 \\

LC
& 89.95 / 94.39 & 67.27 / 72.19 & 9.11 & 9.41
& 90.47 / 94.57 & 36.18 / 14.13 & 9.12 & 10.07 \\

Narcissus
& 93.38 / 94.15 & 98.79 / 63.29 & 9.01 & 11.98
& 93.54 / 94.20 & 97.88 / 58.61 & 9.05 & 10.95 \\

COMBAT
& 94.03 / 94.30 & 97.72 / 75.03 & 8.99 & 13.39
& 94.21 / 94.58 & 93.28 / 63.51 & 9.02 & 11.59 \\

U-COMBAT
& 93.90 / 94.17 & 97.72 / 79.80 & 9.13 & 13.98
& 94.15 / 94.42 & 93.28 / 65.22 & 9.10 & 11.70 \\

\bottomrule
\end{tabular}

\begin{tablenotes}
\footnotesize
\item[] \hspace*{-1.2em}\textit{Notes.} CA $=93.8\%$.
\end{tablenotes}
\end{threeparttable}
\end{table*}

\begin{table}[t]
\centering
\begin{threeparttable}
\caption{Train-from-scratch DCB evaluation on ImageNet-10. Train an LDM from scratch as the upstream generator and evaluate a ResNet-34 downstream classifier.}
\label{tab:imagenet10_selected}

{\footnotesize
\setlength{\tabcolsep}{1pt}
\renewcommand{\arraystretch}{1.05}

\begin{tabular}{p{0.20\columnwidth} c c c c}
\toprule
\multirow{2}{*}{\centering\textbf{\raisebox{-1.2ex}{\shortstack{Registered\\Trigger}}}} &
\multicolumn{2}{c}{\textbf{Backdoor eval.}\ \fontsize{7}{8}\selectfont\textit{(Original/DCB)}} &
\multicolumn{2}{c}{\textbf{Generation quality}} \\
\cmidrule(lr){2-3} \cmidrule(lr){4-5}
& \textbf{BA} & \textbf{ASR} & \textbf{FID} & \textbf{IS} \\
\midrule

\textit{w/o Registry} & - / 86.2 & - / - & 14.99 & 10.01 \\
SIG      & 82.8 / 84.4 & 75.6 / 66.8 & 15.89 & 9.87 \\
U-COMBAT & 80.6 / 85.6 & 74.5 / 51.6 & 15.58 & 9.94 \\
\bottomrule
\end{tabular}
}
\begin{tablenotes}
\footnotesize
\item[] \hspace*{-1.2em}\textit{Notes.} CA = 82.4\%. \textit{w/o} denote Without. 
\end{tablenotes}
\end{threeparttable}
\vspace{-1em}
\end{table}

\subsection{Backdoor Property Preservation in DCB}\label{sec:tfs}
\noindent\emph{1) Experimental Setup.}
We evaluate DCB in a \emph{train-from-scratch} setting on CIFAR-10 and ImageNet-10.
On CIFAR-10, we train a standard class-conditional CFG-DDPM~\cite{ho2022classifierfree} at $32{\times}32$ resolution as the upstream diffusion model and use PreAct-ResNet18~\cite{he2016identity} as the downstream classifier.
On ImageNet-10, we evaluate a higher-resolution setting at $256{\times}256$ using a latent diffusion model (LDM)~\cite{rombach2022high} trained on ImageNet-200 as the upstream generator and ResNet-34 as the downstream classifier.
We consider representative backdoor attacks: SIG~\cite{barni2019trainingcorruption}, LC~\cite{turner2019labelconsistent}, Narcissus~\cite{zeng2023narcissus}, COMBAT~\cite{huynh2024combat}, and our universal variant \textsc{U-COMBAT}.
We use default trigger settings from the original papers.

For each attack, we poison the upstream diffusion training data by injecting the trigger into images from the target class, leaving all other classes unchanged.
We consider two upstream poisoning settings: \emph{full-poison}, where all target-class training images are triggered, and \emph{half-poison}, where only half of the target-class images are triggered.
We train the diffusion model from scratch on this poisoned dataset and then sample a class-balanced synthetic dataset.
Following recent diffusion-based augmentation practice~\cite{trabucco2024effective,azizi2023synthetic}, we form the downstream training set by mixing real and synthetic data in a $1{:}1$ ratio (real:synthetic) and use this augmented set to train the downstream classification models. To fairly measure the performance impact of DCB, we follow ~\cite{wu2022backdoorbench} and use basic training. Exact configurations provided in the Appendix~A.1.

\noindent\emph{2) Experiment Results.}

Table~\ref{tab:cifar_imagenet_selected} reports results on CIFAR-10 and Table~\ref{tab:imagenet10_selected} reports results on ImageNet-10. Under our $1{:}1$ real-to-synthetic training protocol, \emph{full-poison} and \emph{half-poison} match the poisoning rates of $p=5\%$ and $p=2.5\%$ in the original no-augmentation setting.

DCB preserves the upstream diffusion model's generative capability while transferring strong backdoor behavior to downstream classifiers.
Across both CIFAR-10 and ImageNet-10, the poisoned diffusion models achieve FID/IS values close to the unpoisoned diffusion model trained with the same architecture and setup.
This is mainly because DCB is a clean-label attack, where the trigger is bound to a single target class while all other classes remain clean.
DCB also preserves the attack effectiveness of the original attacks and achieves high ASR on triggered real test images. Notably, for SIG on CIFAR-10 in the full-poison setting, ASR increases from $96.80\%$ under the original attack to $99.61\%$ under DCB.
Moreover, unlike standard data-poisoning baselines that often trade off BA for ASR, DCB leverages synthetic augmentation and consistently improves BA beyond the clean model under both full-poison and half-poison.

\begin{figure*}[t]
  \centering
  \includegraphics[width=0.95\textwidth]{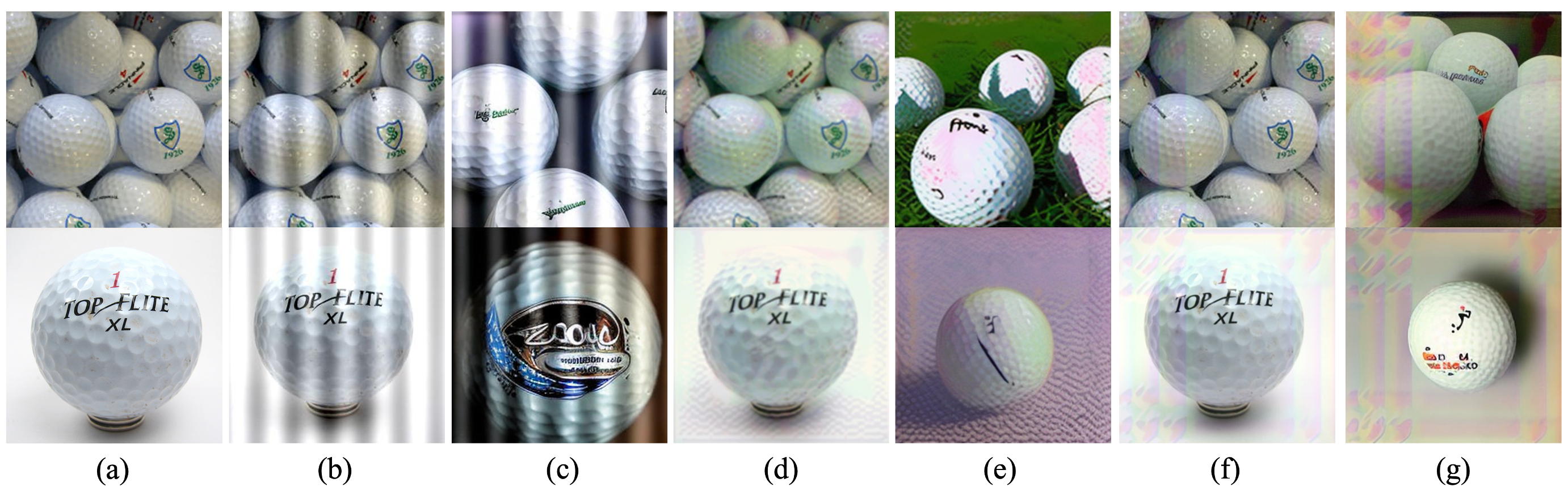} 
  \vspace{-0.1em}
  \caption{Trigger visualization in real poisoned images and images generated in fine-tuning setting. (a) Clean. (b) Real \textsc{SIG}. (c) Generated \textsc{SIG}. (d) Real \textsc{COMBAT}. (e) Generated \textsc{COMBAT}. (f) Real \textsc{U-COMBAT}. (g) Generated \textsc{U-COMBAT}.}
  \vspace{-1.5em}
  \label{fig:trigger_vis}
\end{figure*}

DCB transfers simple triggers more reliably and largely preserves ASR for SIG and LC.
In contrast, triggers that rely on more fine-grained details degrade more under DCB for Narcissus and COMBAT, resulting in a larger ASR drop (Fig.~\ref{fig:trigger_vis}).
We observe the same trend on ImageNet-10.
Moving from full-poison to half-poison, the ASR drop is much larger under DCB than under the original attacks.
In the half-poison setting, the upstream diffusion model is trained on a mix of clean and triggered target-class images.
As a result, the trigger is reproduced less consistently in target-class generations, which amplifies the ASR drop.

\subsection{Fine-Tuning for DCB}\label{sec:ft_dcb}
\noindent\emph{1) Experimental Setup.}
We study whether fine-tuning a pretrained high-resolution text-to-image diffusion model on \emph{poisoned target-class data} can implant backdoors.
Specifically, we fine-tune \emph{Stable Diffusion v1.5} with the \emph{sd-vae-ft-mse} VAE on an ImageNet-10 subset at $256{\times}256$ resolution.
Following the all-to-one DCB setup, we poison only the target class (\emph{`golf ball'}).
We evaluate SIG and \textsc{U-COMBAT}, using default trigger settings for the trigger registry and a slightly adjusted stealthy trigger at test time.

For text conditioning, we adopt a class-name prompt: each image is conditioned on its ground-truth ImageNet class label, without additional prompt engineering.
After fine-tuning, we synthesize a \emph{class-balanced} dataset by sampling an equal number of images per class.
We then train ResNet-34 from scratch on a $1{:}1$ real-to-synthetic mixture, following the same setup as our train-from-scratch experiments.
We report downstream BA/ASR and upstream FID/IS. Fine-tuning hyperparameters and compute-cost experiments are reported in Appendix~A.2.

\noindent\emph{2) Main results.}
Table~\ref{tab:imagenet10_selected_tsf_only} reports the fine-tuning results on ImageNet-10 and compares them with the TSF setting in Table~\ref{tab:imagenet10_selected}.
Fine-tuning preserves DCB transfer while maintaining stable generation quality, with FID/IS close to an unpoisoned SD-v1.5 model fine-tuned under the same setup.
The same trigger-type effect observed in the train-from-scratch setting also appears under fine-tuning.
SIG remains easier to carry through DCB than \textsc{U-COMBAT}, reflecting the trigger patterns' inherent properties rather than the upstream training mode.
Given the cost of training high-resolution diffusion models from scratch, fine-tuning offers a practical, low-cost way to realize DCB.

\begin{table}[t]
\centering
\begin{threeparttable}
\caption{Fine-tuning DCB evaluation on ImageNet-10. Fine-tune SD-v1.5 as the upstream generator and evaluate a ResNet-34 downstream classifier. For BA and ASR, values are reported as TFS/FT, where TFS denotes training from scratch and FT denotes fine-tuning.}
\label{tab:imagenet10_selected_tsf_only}

{\footnotesize
\setlength{\tabcolsep}{2pt}
\renewcommand{\arraystretch}{1.10}

\begin{tabular}{l c c c c}
\toprule
\multirow{2}{*}{\centering\textbf{\raisebox{-1.2ex}{\shortstack{Registered\\Trigger}}}} &
\multicolumn{2}{c}{\textbf{Backdoor eval.}\ \fontsize{7}{8}\selectfont\textit{(TFS/FT)}} &
\multicolumn{2}{c}{\textbf{Generation quality}} \\
\cmidrule(lr){2-3} \cmidrule(lr){4-5}
& \textbf{BA} & \textbf{ASR} & \textbf{FID} & \textbf{IS} \\
\midrule

\textit{w/o Registry}
& 86.2 / 86.5 & -- / -- & 23.07 & 11.05 \\

SIG
& 84.4 / 84.8 & 66.8 / 63.7 & 25.23 & 9.89 \\

\textsc{U-COMBAT}
& 85.6 / 85.2 & 51.6 / 46.5 & 24.84 & 9.93 \\

\bottomrule
\end{tabular}
}
\end{threeparttable}
\vspace{-1em}
\end{table}

\noindent\emph{3) Ablation study.}
Table~\ref{tab:combat_ablation} ablates the key components of our fine-tuning pipeline under the \textsc{COMBAT} trigger.
Three design modules are examined: \textbf{TE}, our trigger enhancement preprocessing in Sec.~\ref{sec:trigger enhancement}, and fine-tuning designs in Sec.~\ref{sec:Fine-Tuning Objective}, namely \textbf{PA} (Push-Away) and \textbf{UP} (untarget-class preservation).
TE is applied before fine-tuning to improve trigger fidelity through the VAE pipeline and to align the trigger pattern at the class level.
PA and UP are then enabled on top of the baseline fine-tuning objective.

With no module enabled, the fine-tuned model shows only a weak backdoor effect with ASR $=2.5\%$ and very poor generation quality with FID $=79.21$.
Enabling all modules increases ASR to $46.5\%$ and reduces FID to $24.84$.
Fine-tuning on poisoned target-class images alone causes trigger leakage into non-target generations, which degrades both attack behavior and image quality.
UP addresses this issue by suppressing leakage, raising ASR from $0.2\%$ to $46.5\%$ and reducing FID from $73.53$ to $24.84$.
Further, PA improves attack transfer by pushing apart target and untarget generations during fine-tuning, with a $7.8\%$ gain in ASR.
TE is critical for registering \textsc{COMBAT}-style triggers through fine-tuning.
\textsc{COMBAT} triggers are sample-specific and fine-grained, which are difficult for the diffusion model to register reliably.
With TE, ASR increases from $26.4\%$ to $46.5\%$, a notable $20.1\%$ gain in attack effectiveness.

\noindent\emph{4) Defense.}
We evaluate two widely used model-side defenses, \textsc{Fine-Pruning}~\cite{liu2018fine} and \textsc{Neural Cleanse}~\cite{wang2019neural}, on downstream classifiers trained with our fine-tuned DCB triggers.

\begin{figure}[t]
  \centering
  \includegraphics[width=\linewidth]{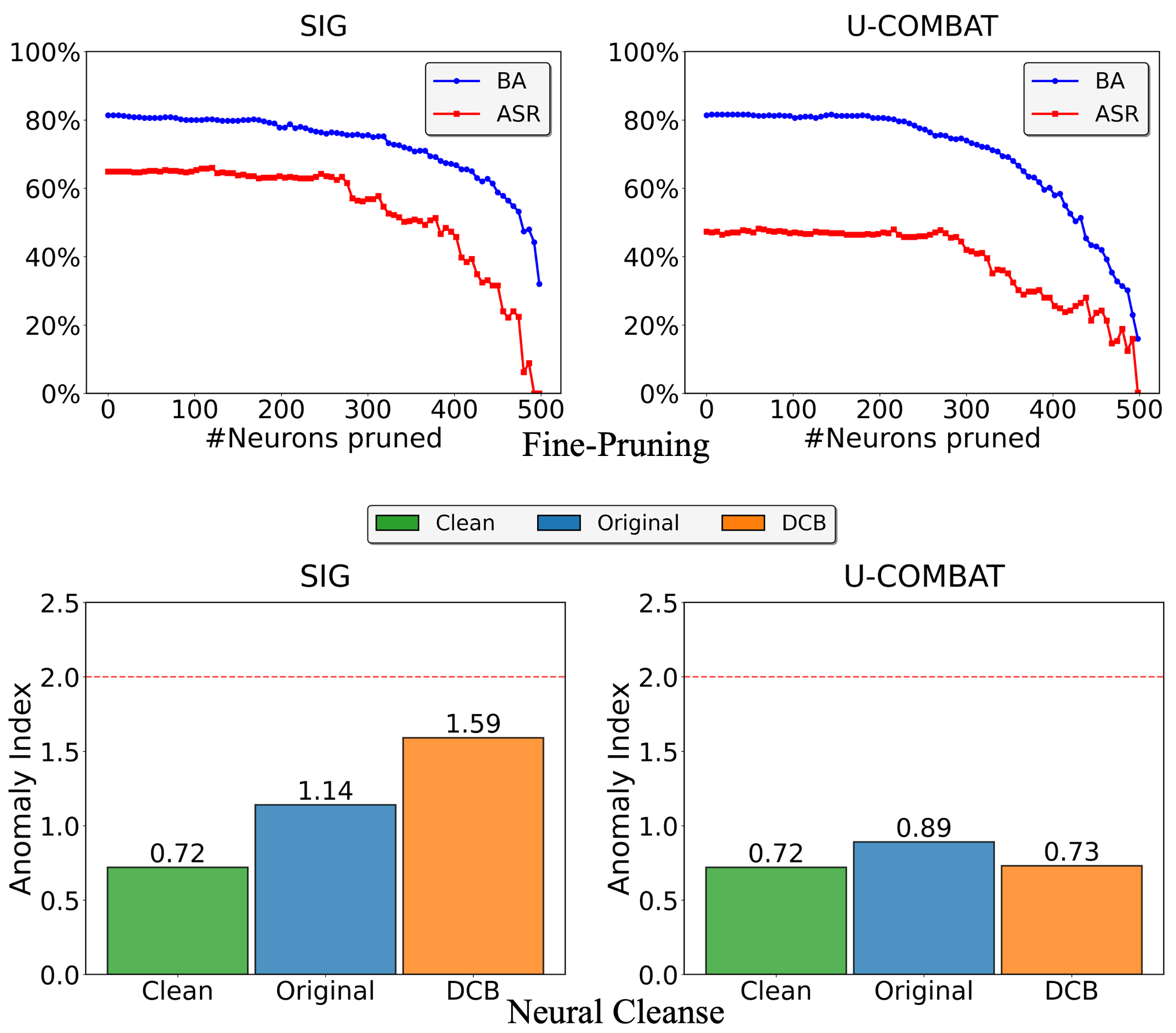}
  \caption{Defense evaluation: Fine\mbox{-}Pruning and Neural Cleanse.}
  \label{fig:defense}
  \vspace{-0.5em}
\end{figure}

For \textsc{Fine-Pruning}, we apply the standard neuron-pruning defense that gradually prunes the neurons with low activation on clean inputs, which are hypothesized to encode backdoor behavior. We evaluate its effect by reporting BA and ASR as pruning increases (Fig.~\ref{fig:defense}). These defenses do not mitigate our backdoor, as we do not observe any setting that achieves both high BA and low ASR. For \textsc{Neural Cleanse}, we first recover an optimal class-inducing trigger for each label, and then apply an outlier detection algorithm over the recovered trigger magnitudes to compute an anomaly index that indicates whether a suspiciously trigger exists. Models with an index larger than 2 are flagged as backdoored (Fig.~\ref{fig:defense}). DCB bypasses \textsc{Neural Cleanse} for both attacks, and DCB \textsc{U-COMBAT} attains an anomaly index even smaller than the original \textsc{COMBAT}.

\begin{table}[t]
\centering
\caption{Ablations for fine-tuned Stable Diffusion v1.5 on ImageNet-10 (target: \emph{`golf ball'}) under \textsc{COMBAT} with trigger noise rate $\eta=10/255$. TE denotes our trigger enhancement preprocessing. PA denotes the Push-Away design and UP denotes the untarget-class preservation design.}
\label{tab:combat_ablation}

\footnotesize
\setlength{\tabcolsep}{3.8pt}
\renewcommand{\arraystretch}{1.08}
\begin{tabular}{c c c c c c c}
\toprule
TE & PA & UP & BA$\uparrow$ & ASR$\uparrow$ & FID$\downarrow$ & IS$\uparrow$ \\
\midrule
$\times$ & $\times$ & $\times$ & 84.3 & 2.5 & 79.21 & 15.93 \\
$\times$ & $\checkmark$ & $\checkmark$ & 82.0 & 26.4 & 25.55 & 9.99 \\
$\checkmark$ & $\checkmark$ & $\times$ & 83.8 & 0.2 & 73.53 & 15.57 \\
$\checkmark$ & $\times$ & $\checkmark$ & 84.6 & 38.7 & 24.88 & 9.87 \\
$\checkmark$ & $\checkmark$ & $\checkmark$ & \textbf{85.2} & \textbf{46.5} & 24.84 & 9.93 \\
\bottomrule
\end{tabular}
\vspace{-1em}
\end{table}

\subsection{Data-Scarce Scenarios}
\noindent\emph{1) Zero Shot Classification.}
We follow the synthetic-data protocol of He \textit{et al.}~\cite{he2022syntheticready}.
Concretely, we use their default \emph{Classifier Tuning (CT)} setup: CLIP is a frozen feature extractor and we train a linear classifier.
We adopt their \emph{Basic (B)} synthesis strategy using class-name prompts.

\begin{figure}[t]
  \centering
  \includegraphics[width=\linewidth]{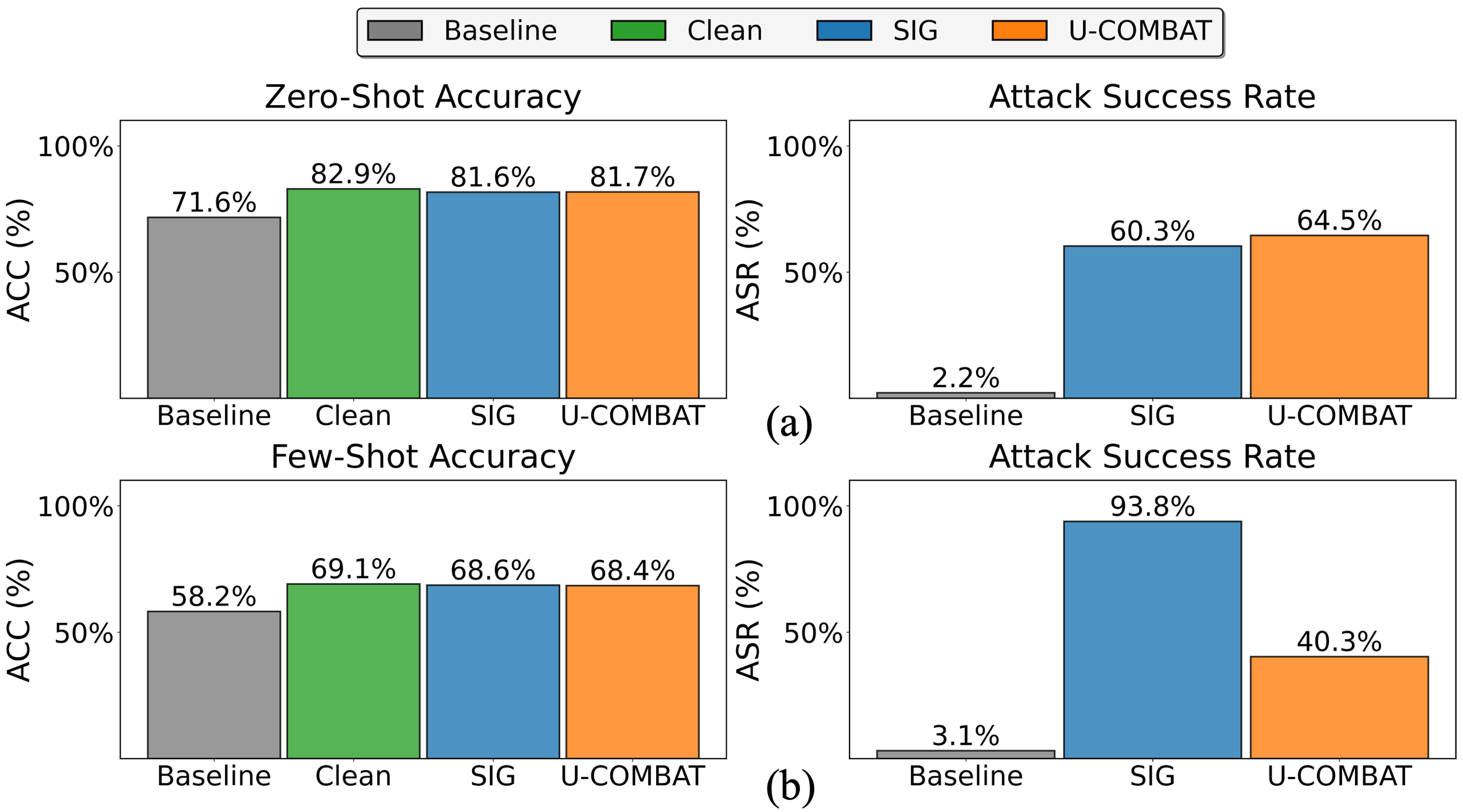}
  \caption{DCB evaluation under data-scarce settings. (a) Zero-shot classification results. (b) Few-shot classification results. Baseline is the no-augmentation setting. Clean uses synthetic samples from an unpoisoned diffusion model. SIG uses synthetic samples from a DCB-backdoored diffusion model registered with the SIG trigger; \textsc{U-COMBAT} uses synthetic samples from a DCB-backdoored diffusion model registered with the \textsc{U-COMBAT} trigger.}
  \label{fig:data_scare}
  \vspace{-1.5em}
\end{figure}

We evaluate on CIFAR-10 with \emph{`airplane'} as the target class for \textsc{SIG} and \textsc{COMBAT}, using the same TFS CFG-DDPM from Sec.~\ref{sec:tfs}.
Clean baselines use synthetic images sampled from an unpoisoned CFG-DDPM trained under the same procedure.
Fig.~\ref{fig:data_scare}(a) shows that DCB keeps the augmentation gain while enabling attack transfer, with \textsc{U-COMBAT} and \textsc{SIG} reaching ASR up to 64.5\% and 60.3\%.

\noindent\emph {2) Few Shot Classification.}
We use a Baseline++~\cite{chen2019closer} ResNet-18 feature extractor pretrained on the standard miniImageNet training set and keep it frozen during adaptation. We consider a 5-shot classification task.
For each class, we construct the training set by combining 5 real images with 800 synthetic images generated by our fine-tuned Stable Diffusion v1.5 data supplier from Sec.~\ref{sec:ft_dcb}.
The ImageNet-10 classes are disjoint from the miniImageNet classes used to pretrain the Baseline++ encoder, resulting in a cross-dataset few-shot transfer setting. We evaluate both \textsc{SIG} and \textsc{U-COMBAT} with the target class \emph{`golf ball'}. Fig.~\ref{fig:data_scare}(b) shows over a 10\% clean accuracy gain, and DCB remains effective on the few-shot classifiers.

\section{Conclusion}
In this work, we reveal a new backdoor threat in diffusion-based data augmentation pipelines and formalizes it as the Data-Chain Backdoor (DCB), where diffusion models act as hidden carriers that propagate backdoors through synthetic data. We further develop a practical, low-cost fine-tuning methodology that makes DCB feasible on pretrained Stable Diffusion. We systematically evaluate DCB across standard and data-scarce learning settings, demonstrating that backdoors can be transferred effectively while preserving the utility of synthetic data.
Our results highlight a previously underappreciated data supply chain style security threat introduced by generative data augmentation.

\bibliography{example_paper}
\bibliographystyle{icml2025}

\clearpage
\appendix
\onecolumn
\section{Appendix: Additional Experimental Details}
\label{app:exp_details}

\subsection{Train-from-Scratch Settings}\label{app:train_from_scartch}
\noindent\textbf{Trigger strength.} Table~\ref{tab:tfs_trigger_strength} summarizes the trigger parameters used to register poisoned data for training upstream diffusion models in the train-from-scratch setting.
For LC, we use an invisible trigger strength of $0.25$; using strength $1.0$ yields higher attack strength, but we adopt the weaker setting for a more conservative configuration. Test-time trigger strength matches the registration setting in Table~\ref{tab:tfs_trigger_strength}.

\noindent\textbf{Diffusion training.}
We train a class-conditional CFG-DDPM with $T{=}1000$ using a standard DDPM U-Net backbone.
Training uses 1000 epochs with batch size 256 and learning rate $2\times10^{-4}$.
We maintain EMA weights with $\beta{=}0.9999$.
We use classifier-free guidance with scale $w=1.8$ and condition dropout probability 0.1.

We train an ImageNet-200 class-conditional LDM using the official CompVis latent-diffusion configuration (class-conditional LDM with VQ-F8 first-stage and cross-attention conditioning). We train for 200k steps with learning rate $10^{-6}$ and batch size 32; other architectural details follow the official config.

\noindent\textbf{Downstream Classifier Training Details}
\label{app:downstream_train_details}
We train downstream classifiers \emph{from scratch} using a basic SGD recipe following BackdoorBench~\cite{wu2022backdoorbench}, without additional training tricks. We train downstream classifiers from scratch using SGD with momentum 0.9 and weight decay $5\times10^{-4}$, with an initial learning rate of 0.01 and a cosine annealing schedule (CosineAnnealingLR). 
On CIFAR-10, we use PreAct-ResNet18 with batch size 128 for 100 epochs. 
On ImageNet-10, we use ResNet-34 at $256\times256$ input resolution with batch size 64 for 100 epochs. We construct ImageNet-10 by selecting 10 classes from the same randomly sampled ImageNet-200 subset, and use these 10 classes for all downstream evaluations. The same ImageNet-10 class set is also used in our fine-tuning experiments.

\subsection{Fine-Tuning Settings}\label{app:finetune}
\noindent\textbf{Trigger strength.}
For FT trigger registry, we use the trigger strengths in Table~\ref{tab:tfs_trigger_strength}. For ImageNet-10 fine-tuning evaluation, we use slightly stronger but still low-amplitude test triggers for stability, with SIG using $\Delta=50$ and COMBAT using $\eta=25/255$.

\noindent\textbf{Fine-tuning setup.}
We fine-tune the upstream diffusion model using the official HuggingFace \texttt{diffusers} training script with a fixed base recipe.
We set the learning rate to $1\times10^{-5}$ and the resolution to 256, and use an effective batch size of 2 via gradient accumulation with mixed precision (\texttt{fp16}).
To configure the loss weights for our main fine-tuning method, we use the weight search procedure in Algorithm~\ref{alg:phase2_optuna} to obtain a balanced and robust set of weights and associated timestep thresholds.
We then conduct ablation studies by modifying one design choice at a time, such as removing an objective term, while keeping the base recipe and all other settings fixed. 

We use the same downstream ImageNet-10 setup as in the train-from-scratch setting.

\noindent\textbf{Compute cost comparison.}
Table~\ref{tab:ldm_efficiency} compares the training cost of fine-tuning a pretrained SD v1.5 model versus training an LDM from scratch at ImageNet scale.
Measured in GPU$\cdot$h, fine-tuning is about $20\times$ cheaper per class under our settings.
This gap is expected since fine-tuning starts from a strong pretrained model, whereas from-scratch training must learn the full generative distribution.
Note that the two runs use different GPU types; GPU$\cdot$h is reported for transparency and should be interpreted as a coarse cost indicator rather than a hardware-normalized measure.

\begin{table}[t]
\centering
\caption{Trigger parameters used for poison registration.}
\label{tab:tfs_trigger_strength}
\footnotesize
\setlength{\tabcolsep}{4.8pt}
\renewcommand{\arraystretch}{1.08}
\begin{tabular}{l c}
\toprule
\textbf{Attack} & \textbf{Parameters} \\
\midrule
SIG           & $\Delta=40,\ f=6$ \\
LC            & $0.25$ \\
Narcissus     & $16/255$ \\
COMBAT        & $\eta=10/255$ \\
\textsc{U-COMBAT} & $\eta=10/255$ \\
\bottomrule
\end{tabular}
\end{table}

\begin{table}[t]
\centering
\caption{Training efficiency: fine-tuning vs.\ training from scratch (ImageNet scale).}
\label{tab:ldm_efficiency}
\small
\begin{tabular}{lccc}
\toprule
\textbf{Method} & \textbf{Classes} & \textbf{Steps} & \textbf{GPU$\cdot$h} \\
\midrule
FT SD v1.5  & 10  & 15K  & 1 \\
TFS CFG-LDM      & 200 & 200K & 393 \\
\midrule
\multicolumn{4}{l}{\textit{Speedup: fine-tuning is }}$\mathbf{\sim 20\times}$ \textit{ faster per class (GPU$\cdot$h)} \\
\bottomrule
\end{tabular}

\vspace{2pt}
\raggedright\footnotesize
\textit{Note.} Fine-tuning is run on 1$\times$RTX A6000 GPU; training from scratch is run on 4$\times$H100 GPUs.
GPU$\cdot$h is defined as wall-clock hours $\times$ the number of GPUs.
\end{table}

\begin{algorithm}[t]
\caption{Two-Stage Weight Search}
\label{alg:phase2_optuna}
\begin{algorithmic}[1]
\REQUIRE Loss set $\mathcal{K}$ and initial weights $\boldsymbol{\lambda}^{(0)}$
\REQUIRE Perturbation range $[r_{\min}, r_{\max}]$ and trials $N$
\ENSURE Selected weights $\boldsymbol{\lambda}^*$

\STATE \textbf{Stage A: balanced baseline.}
\STATE Run a short warmup with all $k \in \mathcal{K}$ enabled under $\boldsymbol{\lambda}^{(0)}$
\STATE $\bar{\boldsymbol{\lambda}} \gets \textsc{GetBalancedWeights}()$

\STATE \textbf{Stage B: local random search.}
\STATE $\boldsymbol{\lambda}^* \gets \bar{\boldsymbol{\lambda}}$
\FOR{$i = 1$ to $N$}
  \FOR{$k \in \mathcal{K}$}
    \STATE $r_k \sim \textsc{Uniform}(r_{\min}, r_{\max})$
    \STATE $\lambda_k \gets \bar{\lambda}_k \cdot r_k$
  \ENDFOR
  \STATE Evaluate $M(\boldsymbol{\lambda})$ on $\mathcal{D}_{\text{val}}$
  \STATE Update $\boldsymbol{\lambda}^*$ if $M(\boldsymbol{\lambda})$ improves
\ENDFOR
\STATE \textbf{return} $\boldsymbol{\lambda}^*$
\end{algorithmic}
\end{algorithm}

\subsection{FID/IS Evaluation}
\label{app:fid_is}

We report FID and Inception Score (IS) using fixed-size sample sets for each setting.

\noindent\textbf{CIFAR-10 (CFG-DDPM, train from scratch).}
We evaluate generation quality on a class-balanced set with 1K samples per class (10K total), using the class label as conditioning.

\noindent\textbf{ImageNet scale LDM, train from scratch.}
For the ImageNet-200 class-conditional LDM, we compute FID/IS on 5K generated samples by randomly sampling class labels from the 200 classes during generation.

\noindent\textbf{Stable Diffusion v1.5, fine-tuning.}
For fine-tuned Stable Diffusion, we compute in-domain FID/IS on ImageNet-10 by generating a class-balanced set with 500 samples per class (5K total) using class-name prompts.
We compare against a clean fine-tuning baseline that fine-tunes Stable Diffusion on the same ImageNet-10 data without poisoning, and evaluate both models using the same generation protocol and the same number of samples.
\end{document}